\documentstyle[12pt,psfig]{article}
\begin{document} 

\title{Final state interaction in near-threshold meson
production\thanks{Supported 
by Forschungszentrum J\"ulich}}
\author{A. Sibirtsev \\
Institut f\"ur Theoretische Physik, Universit\"at Giessen \\
D-35392 Giessen, Germany}
\date { }
\maketitle

\phantom{a}\vspace*{+1mm}
It was predicted by Watson and Migdal~\cite{Migdal} that 
close to  threshold the energy dependence of the cross section 
is dominated by the Final State Interaction (FSI). 
Within the Watson-Migdal approximation
the transition amplitude is factorized in terms of the production
and FSI amplitudes. Further study  
were performed by Gottfried and Jackson~\cite{Gottfried} with
introduction the absorptive correction to the Born amplitude
due to the initial and FSI. 

Here we address the question  what information, other
than the FSI parameters, one can extract from 
data on near-threshold meson production.
 
\begin{figure}[h]
\phantom{a}\vspace*{-5mm}\hspace*{-0.5cm}
\vspace*{2mm}
\psfig{file=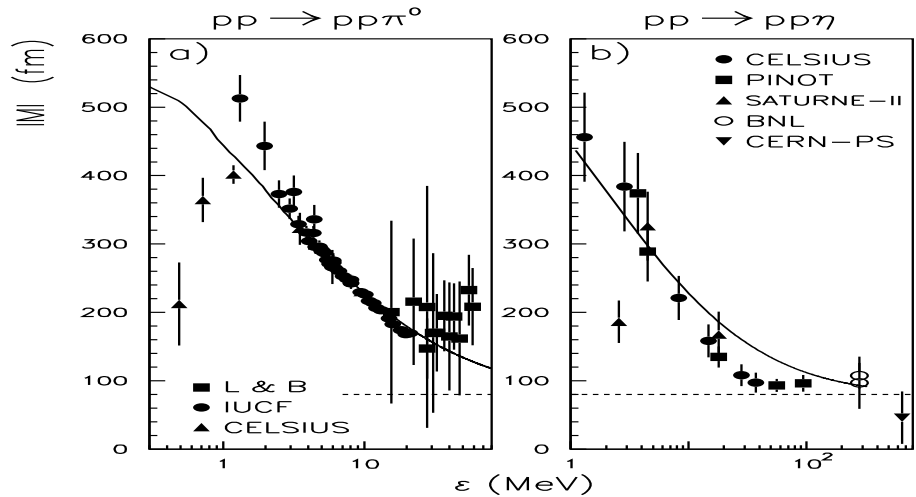,height=7cm,width=14cm}
\caption[]{The amplitudes for the $pp \to pp \pi^0$ (a) and
$pp \to pp \eta$ (b) reactions.
The symbols show the results extracted from the 
experimental data~\protect\cite{LB,pion,eta}.
The solid line shows the averaged $pp$ scattering amplitude.}
\label{conf2}
\end{figure}

As was first proposed by Gell-Mann and
Watson~\cite{GellMann} the near-threshold  energy dependence 
of the $pp \to pp \pi^0$  cross section is well 
reproduced by the Phase-Space (PS)
basis and the FSI. Within the nonrelativistic limit the PS 
for $n$-particles is proportional to $\epsilon^{(3n-5)/2}$, 
where $\epsilon$ stands
for the excess energy and equals to the difference between the invariant
collision energy and the total mass of the produced particles.
At the range $0\le \epsilon \le 100$~MeV the cross section
for 3-particle production might increase 4 orders of magnitude
due to the PS only. Let us to analyze not the cross section 
itself but the average reaction amplitude given as
\begin{equation} 
|M| = \left( F \sigma /R_3 \right)^{1/2}
\end{equation}
where $F$ is the flux factor and $R_3$ is the three-body PS.
Fig.~\ref{conf2}a) shows the $|M|$ evaluated from the  
$pp \to pp \pi^0$ data~\cite{LB,pion}.
If the prediction~\cite{GellMann} is true, the data should
indicate the FSI between the final protons. Taking the production
amplitude as a constant~\cite{Sibirtsev}, the $|M|$ is factorized in
terms of this constant and the $pp$ scattering amplitude
$T_{pp}$.
Solid line in Fig.~\ref{conf2} shows the  $T_{pp}$
from the Nijmegen-93 model~\cite{Nijmegen}  averaged over the
available PS. The discrepancy at low $\epsilon $ is
due to the repulsive Coulomb interaction not
incorporated in the present calculations. 

Similar situation holds for $pp \to pp \eta $ reaction. 
Fig.~\ref{conf2}b) shows the reaction amplitude extracted from the 
data~\cite{LB,eta} together with the average 
$T_{pp}$~\cite{Sibirtsev}. Both $pp \to pp \pi^0$ and
$pp \to pp \eta$ reactions indicate the same strength of
the FSI at the same range of the excess energy. Moreover,
Fig.~\ref{conf2} obviously illustrates the validity of the
Watson-Migdal approximation.

The aim of the near-threshold meson production experiments
is not to measure the $pp$ scattering amplitude, which
can be determined more precisely through the partial waves 
analysis. The crucial measure for meson
production is the production amplitude, which is shown in 
Fig.~\ref{conf2} with the dashed line. One possible
and almost model independent\footnote{in sense of the
Watson-Migdal approximation} way to evaluate the
production mechanism from the  data
is to make the FSI corrections based on the known scattering
amplitude similar to the procedure we performed for
$pp \to pp \pi^0$ and $pp \to pp \eta$ reactions. Obviously
it is necessary to measure the reaction at  wide range of
$0 \le \epsilon \le 100$~MeV in order to escape ambiguity of
the analysis.

\begin{figure}[h]
\phantom{a}\vspace*{-5mm}\hspace*{-0.5cm}
\hspace*{-0.5cm}
\vspace*{2mm}
\psfig{file=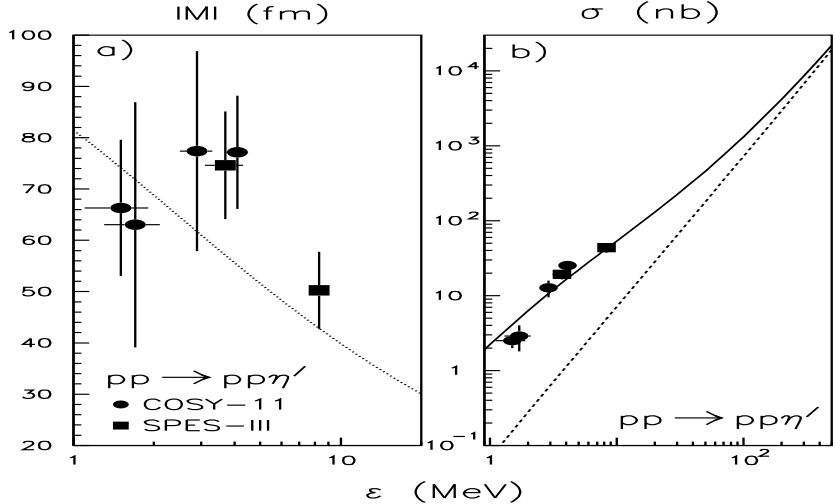,width=14cm,height=7cm}
\caption[]{The amplitude (a) and cross section (b) 
for the $pp \to pp \eta'$  reaction.
The experimental data are from Ref.~\protect\cite{etap}. 
The dotted line in a) shows the $T_{pp}$ averaged over the
PS, the solid line  in b) shows the one-pion exchange calculations 
with FSI, while the dashed line in b) -without FSI.}
\label{conf4}
\end{figure}

As an example one can look recent data~\cite{etap} on 
$pp \to pp \eta'$ reaction. The average reaction amplitude is
shown in Fig.~\ref{conf4}a). Neglecting the FSI one can fit the 
data with the constant $|M|\simeq 70$~fm and motivate that the
cross section is reproduced by the PS alone. However, from
near-threshold $\pi^0$ and $\eta$-production we learned
the FSI dominance, which is indicated with the dotted line in 
Fig.~\ref{conf4}a). 

Fig.~\ref{conf4}b) shows  one-pion exchange~\cite{Sibirtsev} 
calculations with (solid line) and  
without (dashed line) FSI between the final protons. The
model with FSI quite reasonably reproduce the data, however 
we need more data at the range $\epsilon \simeq 100$~MeV
for crucial verification.

\end{document}